\documentclass{pasj00}

\begin{document}
\SetRunningHead{Tsujimoto, M. et al.}{VLA Observation of a Protostar in OMC-3}
\Received{}
\Accepted{}

\title{A High-Resolution Very Large Array Observation of a Protostar in OMC-3:
Shock-induced X-ray Emission by a Protostellar Jet}
\author{
Masahiro \textsc{Tsujimoto},\altaffilmark{1}
Katsuji \textsc{Koyama},\altaffilmark{2}
Naoto \textsc{Kobayashi},\altaffilmark{3}
Masao \textsc{Saito},\altaffilmark{4}
Yohko \textsc{Tsuboi},\altaffilmark{5}
\and
Claire J. \textsc{Chandler}\altaffilmark{6}
}
\altaffiltext{1}{Department of Astronomy \& Astrophysics, Pennsylvania State University,\\
525 Davey Laboratory, University Park, PA 16802, USA}
\email{tsujimot@astro.psu.edu}
\altaffiltext{2}{Department of Physics, Graduate School of Science, Kyoto University,\\
Kitashirakawa Oiwake-cho, Sakyo-ku, Kyoto, 606-8502}
\altaffiltext{3}{Institute of Astronomy, University of Tokyo, 
2-21-1 Osawa, Mitaka, Tokyo, 181-0015}
\altaffiltext{4}{National Astronomical Observatory of Japan, 
2-21-1 Osawa, Mitaka, Tokyo, 181-0015}
\altaffiltext{5}{Department of Physics, Faculty of Science and Engineering, Chuo University,\\
1-13-27 Kasuga , Bunkyo-ku, Tokyo, 112-8551}
\altaffiltext{6}{National Radio Astronomy Observatory, 1003 Lopezville Road, Socorro, NM 87801, USA}

\maketitle

\begin{abstract}
 Using the Very Large Array (VLA) in the A-configuration, we have obtained a high-resolution
 3.6\,cm map of a hard X-ray source detected by the Chandra X-ray Observatory
 in a protostellar clump in Orion molecular cloud 3. Two radio continuum sources
 were detected in the vicinity of the X-ray source, both of which have NIR
 counterparts. We conclude that these VLA sources are free-free emission
 produced by shocks in protostellar jets from the NIR class~I protostars. Using the
 centimeter data, we determined the power and orientation of the protostellar jets.
 The center position of the X-ray emission was found to be
 $\sim$\,1--2\arcsec\ offset from the exciting sources of the jets, and the displacement
 is in the direction of the jets and molecular outflows. We discuss the nature of the
 X-ray emission as the shock-excited plasma at the shock front where the jet propagates
 through interstellar medium at a speed of $\sim$\,1000\,km\,s$^{-1}$.
\end{abstract}

\section{Introduction}
Since the discovery of X-ray emission from star-forming regions (SFRs) using the
Einstein Observatory in the 1980's \citep{feigelson81,montmerle83}, low-mass young
stellar objects (YSOs) have been known to be strong X-ray emitters. Subsequent X-ray
observatories further revealed that almost all low-mass YSOs in the class~I---III
evolutionally stages emit X-rays and the emission originates from optically-thin thermal plasma with
fast rises and slow decays in flux. These lead to the general consensus that
X-rays from low-mass YSOs are attributable to high temperature
($T$=10$^{6}$--10$^{7}$\,K) plasma maintained by occasional flares that are triggered by
magnetic reconnections (see a review by \cite{feigelson99}).

Recent long-exposure observations by the Chandra X-ray Observatory are beginning to
reveal new X-ray-emitting phenomena in SFRs with its unprecedented sensitivity and
spatial resolution. These include the X-ray detection from a Herbig-Haro object
\citep{pravdo01} and the discovery of diffuse X-ray emission from high-mass SFRs (e.g.,
\cite{townsley03}).

\medskip

We conducted a $\sim$\,100\,ks observation on Orion molecular clouds 2 and 3 (OMC-2/3;
$D$=450\,pc) with the Advanced CCD Imaging Spectrometer (ACIS) onboard Chandra
\citep{tsuboi01,tsujimoto02a} in conjunction with deep near-infrared (NIR) follow-up
imaging using the University of Hawaii 88 inch telescope \citep{tsujimoto03}. Among 385
X-ray detections, we found that $\sim$\,72\% sources have a NIR counterpart while the
remaining sources have none. Based on the X-ray spectral and temporal properties as well
as the spatial distribution and the \textit{K}-band luminosity function of these
sources, we concluded that most of the NIR-identified and NIR-unidentified X-ray sources
are YSOs and background active galactic nuclei, respectively.

Among the NIR-unidentified X-ray sources, however, a few coincide spatially with the
string of protostellar clumps in OMC-3 (MMS\,1--10) seen in 1.3\,mm continuum
emission \citep{chini97}. The hard X-ray emission (source No.\,8 in \cite{tsuboi01}; we
hereafter call this source TKH\,8) detected at the protostellar clump MMS\,2 is one of
these peculiar X-ray sources. Its associations with the protostellar clump, with molecular
outflows in CO, HCO$^{+}$ \citep{aso00,williams03}, and shock-excited H$_{2}$
\citep{tsujimoto02b} emission lines, and with protostellar jets in centimeter continuum
\citep{reipurth99}, make this source unlikely to be a background source. It is also
unlikely that this X-ray emission is from the magnetic activity of a low-mass
class~I--III objects because the X-ray emission is centered at a position that is
significantly ($\sim$\,1--2\arcsec) offset from the closest NIR sources (IRS\,3 and
IRS\,5; \cite{tsujimoto02b}). These results lead \citet{tsuboi01} and \citet{tsujimoto02b} to
speculate that this X-ray emission is from a hidden class~0 protostar, which is
invisible by definition at NIR wavelengths \citep{barsony94}.

\medskip
We have carried out the highest-resolution centimeter imaging observation yet of TKH\,8 using the
Very Large Array (VLA) to investigate further the nature of this peculiar source. Given
the fact that most protostars show free-free emission originating from protostellar
jets \citep{rodriguez95,andre96}, long-baseline interferometer imaging at centimeter
wavelengths is the most accurate method of determining the position of jet-exciting
sources, and the strength and orientation of the jet at its root. A D-configuration
observation has already detected a 3.6\,cm source at TKH\,8 \citep{reipurth99}, although
its coarse spatial resolution is insufficient to match the Chandra image. In this
paper, we report the result of our A-configuration VLA observation of TKH\,8, based on
which we present another interpretation of the X-ray emission.

\section{Observation \& Result}
We observed TKH\,8 with the VLA of the National Radio Astronomy Observatory (NRAO) on
February 11, 2002. We used the A-configuration to achieve sufficient spatial resolution
to match that of the Chandra image. A 3.6\,cm map was obtained with an
integration time of $\sim$\,3.5 hours and the phase center at
R.\,A.$=$\timeform{05h35m18.3s}, Decl.$=$\timeform{-05D00'33''} (J2000.0). The map is sensitive to
structure smaller than $\sim$\,2\arcsec\ with an angular resolution of
$\sim$\,0\farcs1. This enables us to have a much finer view of the X-ray emission and
its vicinity compared to the prior D-configuration observation, which had a resolution of
$\sim$\,8\arcsec\ \citep{reipurth99}. 3C\,48 (3.25\,Jy) and 0541$-$056 (0.98\,Jy) were used as
the flux and phase calibrators, respectively.

\begin{figure}
 \begin{center}
  \FigureFile(80mm,80mm){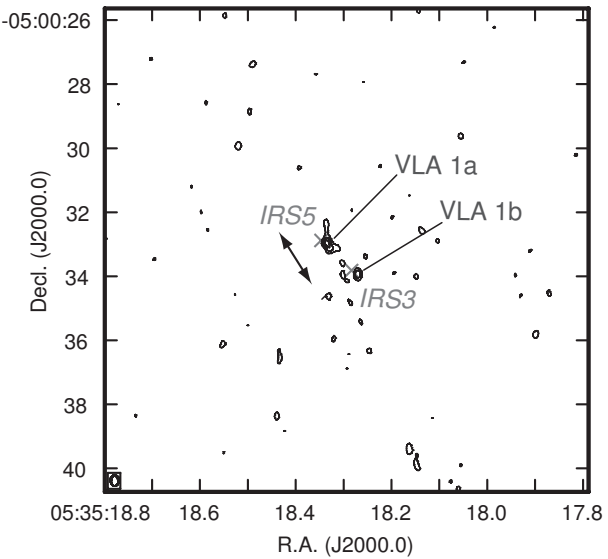}
 \end{center}
 \caption{The 3.6\,cm intensity map (contours) with the position of the NIR sources
 (crosses). The position accuracy of the VLA sources are $\sim$\,0\farcs02, while that
 of the NIR sources are $\sim$\,0\farcs1. Contour levels are 3--9\,$\sigma$ with a step
 of 3\,$\sigma$, where the background noise is $\sim$\,7.7\,$\mu$Jy\,beam$^{-1}$. The
 synthesized beam size is at the bottom left. The position angle of VLA\,1a is shown
 with the arrow.}\label{fg:f1}
\end{figure}

Data reduction, calibration and analysis were performed using the Astronomical
Image Processing System (AIPS). The naturally-weighted map is shown in
figure~\ref{fg:f1}. We detected two sources (VLA\,1a and VLA\,1b) above the 3\,$\sigma$
level, for which we derived the position and the flux density (table \ref{tb:t1}). These
two sources were not resolved in the prior D-configuration observation and were named
altogether as VLA\,1 \citep{reipurth99}. We detected $\sim$\,70\% of the total flux
measured in the D configuration, indicating that we are missing some extended emission
associated with VLA\,1. VLA\,1a is slightly extended, so we also determined the length
of the major (minor) axis and position angle to be 0\farcs40 (0\farcs09) and
32.2~degree, respectively. The direction of the elongation is shown with the arrow in
figure~\ref{fg:f1}.

\begin{table*}
 \begin{center}
  \caption{Source list.}\label{tb:t1}
  \begin{tabular}{ccccc}
   \hline\hline
   source & R.A. & Decl. & flux\footnotemark[$*$] & NIR \\
   ID     & (J2000.0) & (J2000.0) & ($\mu$Jy) & counterpart\footnotemark[$\dagger$]\\
   \hline
   VLA\,1a & \timeform{05h35m18.335s} & \timeform{-05D00'32.97''} & 128$\pm$18 & IRS\,5\\
   VLA\,1b & \timeform{05h35m18.271s} & \timeform{-05D00'33.93''} & \phantom{0}54$\pm$11 & IRS\,3\\
   \hline
   \multicolumn{4}{@{}l@{}}{\hbox to 0pt{\parbox{85mm}{
   \footnotesize
   \footnotemark[$*$] The flux density is corrected for the primary beam response.
   \par\noindent
   \footnotemark[$\dagger$] The nomenclature follows \citet{tsujimoto02b}.
   }\hss}}
  \end{tabular}
 \end{center}
\end{table*}

\section{Discussion}
\subsection{The Origin of the Centimeter Emission}
VLA\,1a and VLA\,1b are very close to the NIR sources IRS\,5 and IRS\,3, respectively
\citep{tsujimoto02b}. The apparent separation between IRS\,5  and VLA\,1a
($\sim$\,0\farcs1) and between IRS\,3 and VLA\,1b ($\sim$\,0\farcs07) are smaller than
the NIR position uncertainty of $\sim$\,0\farcs1. Thus, the NIR and radio sources are
most likely coincident. Both IRS\,3 and IRS\,5 can be classified as class~I protostars
from their NIR excess and high extinction of $A_{V}>$50~mag using the
(\textit{H}--\textit{K})/(\textit{K}--\textit{L}$^{\prime}$) color-color diagram
(figure~3b in \cite{tsujimoto02b}). The excess emission was also confirmed at the
mid-infrared bands, securing the class~I nature of these sources \citep{nielbock03}.

\medskip

For the following four reasons, we conclude that both VLA\,1a and VLA\,1b are free-free
emission from H$_{\rm{II}}$ regions ionized by the UV radiation from the shock front
produced by a protostellar jet propagating through ambient matter \citep{curiel87}.

First, these centimeter sources are accompanied by class~I protostars (IRS\,3 and
IRS\,5). Embedded protostars are frequently associated with centimeter
emission. Detailed studies indicate that most of these radio sources, if not all, are jet-induced
free-free emission \citep{anglada96}.

Second, IRS\,3 and IRS\,5 have the NIR magnitudes and colors of \textit{J}$_{0} >$
11.3~mag and \textit{J}--\textit{H} $>$ 1.6~mag, and \textit{J}$_{0} >$ 11.3~mag and
\textit{J}--\textit{H} $>$ 3.8~mag, respectively \citep{tsujimoto02b}. This indicates
that both sources have a mass less than 2\,$M_{\odot}$, which rules out the possibility
that the centimeter emission is from H$_{\rm{II}}$ regions generated by stellar UV
photons.

\begin{figure}
 \begin{center}
  \FigureFile(80mm,80mm){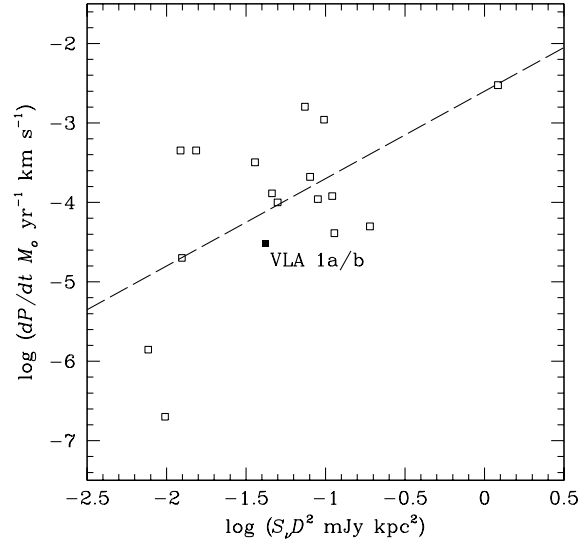}
 \end{center}
 \caption{Relation between $S_{\nu}D^{2}$ and $dP/dt$. Open squares are from
 \citet{anglada92}, who derived an empirical relation for these sources (dashed
 line). VLA\,1a/b (filled square) is consistent with this relation.}\label{fg:f2}
\end{figure}

Third, the flux density multiplied by the square of the distance to the source from an
observer ($S_{\nu}D^{2}$) and the momentum rate in the outflow ($dP/dt$) of VLA\,1a and
VLA\,1b fit well with a known empirical relation \citep{anglada92} and theoretical
understanding \citep{curiel87} of the jet-induced model. Figure~\ref{fg:f2} shows the
relation between $S_{\nu}D^{2}$ and $dP/dt$ for 16 embedded objects, where we added
VLA\,1a/b with $S_{\nu}D^{2}=0.182\times0.45^{2}$\,mJy\,kpc$^{2}$ (this work) and
$dP/dt=3\times10^{-5}$\,$M_{\odot}$\,yr$^{-1}$\,km\,s$^{-1}$ \citep{aso00}. Here, the
limited spatial resolution of the HCO$^{+}$ and CO observations in \citet{aso00} did not
resolve which of the three 1.3\,mm clumps (MMS\,2--MMS\,4; \cite{chini97}) is responsible
for the molecular outflow. However, VLA\,1a and VLA\,1b in MMS\,2 are the only 3.6\,cm
sources that are associated with MMS\,2--MMS\,4 (this work; \cite{reipurth99}). We
therefore safely assumed that the $dP/dt$ value determined for this molecular outflow
represents the sum of the momentum rate from VLA\,1a and VLA\,1b. Similarly, we summed
the flux density of VLA\,1a and VLA\,1b for the $S_{\nu}$ value.

\begin{figure}
 \begin{center}
  \FigureFile(80mm,52mm){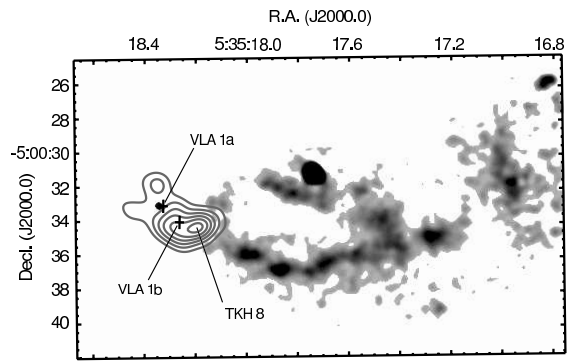}
 \end{center}
 \caption{The H$_{2}$ intensity map (gray scale) with the position of the 3.6\,cm
 sources (pluses) and the hard X-ray intensity (contours).}\label{fg:f3}
\end{figure}

Fourth, in case of VLA\,1a, the source is elongated (the arrow in figure~\ref{fg:f1}) in
the direction of the global outflow seen in the H$_{2}~v=1-0$~S(1)-band
(\cite{tsujimoto02b}; the main stream emanating from VLA 1a/1b in figure~\ref{fg:f3}),
which is characteristic of free-free centimeter emission induced by jets
\citep{anglada96}.

\subsection{The Origin of the Hard X-ray Emission}\label{sec:s3-2}
The poor photon statistics of the hard X-ray source TKH\,8 do not allow us to examine
whether this is an extended source. However, with the assumption that it is a point-like
source, the X-ray emission is significantly offset from the class~I protostars (IRS\,3
and IRS\,5; \cite{tsujimoto02b}). Our high-resolution radio map enables us to confirm that the
displacement is in the direction of the jet and outflow. We discuss the origin of this
X-ray emission together with the centimeter emission based on a jet-induced scenario.

\begin{figure}
 \begin{center}
  \FigureFile(80mm,48mm){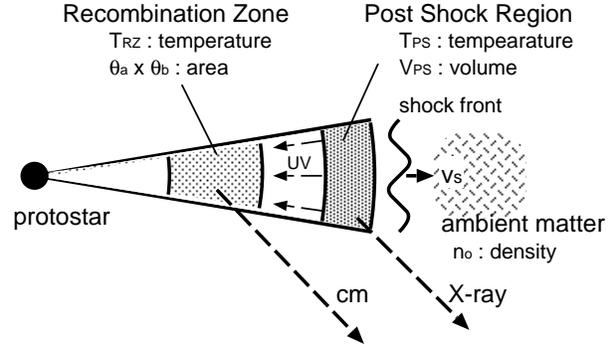}
 \end{center}
 \caption{Schematic view of the centimeter and X-ray emission at the shock
 front.}\label{fg:f4}
\end{figure}

Figure~\ref{fg:f4} shows a schematic view, where the hard X-rays are emitted from the
post shock (PS) region, while the centimeter emission is from the recombination zone
(RZ) behind the shock front. The RZ is maintained by the continuous ionization by UV photons
from the PS region.

In PS, the temperature ($T_{\rm PS}$) and the density ($n_{\rm PS}$) are expressed as
\begin{equation}
 T_{\rm PS}=1.5\times10^{5}\left(\frac{v_{\rm s}}{100\,\rm{km\,s}^{-1}}\right)^2~[\rm K],\label{eq:e1}
\end{equation}
\begin{equation}
 n_{\rm PS}= 4n_{0}~[{\rm cm}^{-3}],\label{eq:e2}
\end{equation}
where $v_{\rm s}$ is the velocity of the shock produced by the collision of the jet into
ambient matter with density $n_{0}$ \citep{raga02}. When $v_{\rm s}$ is large
enough, we can expect to have X-ray emission from PS. Assuming that light elements
are fully ionized in PS, the emission measure ($EM$) is given with the electron density
($n_{\rm PS}$) and the volume ($V_{\rm PS}$) of PS as
\begin{equation}
 EM = n_{\rm PS}^{2}V_{\rm PS}~[\rm{cm}^{-3}].\label{eq:e3}
\end{equation}
We fitted the X-ray spectrum of TKH\,8 with an optically-thin thermal plasma model with the
amount of interstellar absorption, plasma temperature, and emission measure as free parameters,
obtaining an acceptable fit with $N_{\rm H}=12.7 \times 10^{22}$\,cm$^{-2}$,
$k_{\rm{B}}T_{\rm PS}=3.05$\,keV, and $EM=2.1\times10^{53}$\,cm$^{-3}$. Readers should
note, however, that these parameters are not well constrained due to the poor X-ray
photon statistics of this source, which can bring a large uncertainty in the following
calculations. Using the best-fit values, we derive $v_{\rm
s}=1.5\times10^{3}$\,km\,s$^{-1}$ and $n_{0}=5.8\times10^{2}$\,cm$^{-3}$. Here, we
assumed that PS is a cube at a distance of 450\,pc with the length of 0\farcs5 ($=$ the
pixel scale of ACIS).

\medskip

The radio observations give us another constraint on the values of $v_{\rm s}$ and
$n_{0}$ as shown below. In RZ, the centimeter intensity ($S_{\nu}$) is given by
\begin{eqnarray}
 \left(\frac{S_{\nu}}{\rm{mJy}}\right) &=& 1.42 \times 10^{2} 
  \left(\frac{\theta_{a}\theta_{b}}{\rm{arcsec^{2}}}\right)
  \left(\frac{\nu}{5\,\rm{GHz}}\right)^{2}
  \nonumber\\ && \times
  \left(\frac{T_{\rm{RZ}}}{10^{4}\,\rm{K}}\right)
  \left\{1-\exp{(-\tau_{\nu})}\right\},\label{eq:4}
\end{eqnarray}
where $\theta_{a}$ and $\theta_{b}$ are the axis lengths of the extended centimeter
source and $T_{\rm{RZ}}$ is the temperature in RZ \citep{curiel93}. The term
$1-\exp{(-\tau_{\nu})}$ approaches $\tau_{\nu}$ in the optically-thin limit, which is
expressed in terms of the shock parameters \citep{curiel89} as
\begin{eqnarray}
 \tau_{\nu} &=& 1.55\times10^{-7}\left(\frac{n_{0}}{1\,\rm{cm}^{-3}}\right)\left(\frac{v_{\rm s}}{100\,\rm{km\,s}^{-1}}\right)^{1.68}
  \nonumber\\ && \times 
  \left(\frac{\nu}{5\,\rm{GHz}}\right)^{-2.1}\left(\frac{T_{\rm RZ}}{10^{4}\,\rm{K}}\right)^{-0.55}.\label{eq:e5}
\end{eqnarray}
By substituting the observed values for VLA\,1a ($\nu=$8.3\,GHz, $S_{\nu}=0.128$\,mJy,
$\theta_{\rm{a}}=$0\farcs40, and $\theta_{\rm{b}}=$0\farcs09) as typical values and
assuming that $T_{\rm RZ}=10^{4}$\,K, we obtain
\begin{equation}
 \left(\frac{n_{0}}{1\,\rm{cm}^{-3}}\right)\left(\frac{v_{\rm s}}{100\,\rm{km}\,\rm{s}^{-1}}\right)^{1.68}=1.7\times10^{5}.\label{eq:e6}
\end{equation}

From the X-ray observations, we can independently derive that
\begin{equation}
 \left(\frac{n_{0}}{1\,\rm{cm}^{-3}}\right)\left(\frac{v_{\rm s}}{100\,\rm{km}\,\rm{s}^{-1}}\right)^{1.68}=5.7\times10^{4}.\label{eq:e7}
\end{equation}
These values are in a good agreement with each other within a factor of a few,
supporting the protostellar jet scenario for the origin of the X-ray and centimeter
emission.

\subsection{A Similar Example}
Similar X-ray emission was recently reported by \citet{bally03} from L1551 IRS\,5, in
which an X-ray source detected by Chandra is located at the base of the
protostellar jet. They argued three possibilities as the origin of the X-rays, including
X-rays from fast shocks. Applying our scenario to this source using the X-ray
observables given in \citet{bally03} and the centimeter observables in \citet{cohen82},
we found that the discussion in Sect.\,\ref{sec:s3-2} also holds for this source with
$v_{s}\sim$\,500\,km\,s$^{-1}$. This velocity is reinforced by \citet{pyo02}, who
present NIR echelle spectroscopy results on the [Fe$_{\rm{II}}$] outflow from this
source, and indicate the existence of a high velocity component reaching
400--500\,km\,s$^{-1}$. 

A recent series of works using [Fe$_{\rm{II}}$] spectroscopy detected a highly-collimated
high velocity component exceeding $\sim$\,200\,km\,s$^{-1}$ in protostar outflows,
including DG Tau (250--300\,km\,s$^{-1}$; \cite{pyo03}) and HL Tau
($\sim$\,250\,km\,s$^{-1}$; T-S. Pyo, private communication). In optical jet
observations, FS Tau B was measured both in radial velocity and proper motion,
giving a jet velocity of $\sim$\,400\,km\,s$^{-1}$ \citep{eisloeffel98}. Future X-ray
observations of these systems with active jets are needed, to enrich the sample
and examine the correlation between X-ray and centimeter observables, in order to test
our idea.

\subsection{Implications}
Protostellar jets at a speed of 500--1000\,km\,s$^{-1}$ may sound extreme. However,
such a high velocity should not be surprising if we consider that even a Herbig-Haro
object, which is far away from its powering source, is emitting X-rays induced by a
$\sim$\,200\,km\,s$^{-1}$ shock \citep{pravdo01}, and also that solar coronal mass
ejections are frequently observed to be propagating at $\gtrsim$\,1000\,km\,s$^{-1}$ (e.g.,
\cite{gallagher03}). The lack of detection of 500--1000\,km\,s$^{-1}$ jets so far does not
prove their non-existence but simply means we have had no tools to measure them.

If this picture is established, we will obtain a strong method to derive the speed of
highest velocity component of protostellar jets from X-ray imaging-spectroscopy using
equation (\ref{eq:e1}), with no dependence on the inclination angle unlike proper motion
or Doppler shift measurements. We consider more observations of similar sources will
boost our understanding of this phenomenon, and eventually, of forming stars.

\section{Summary}
\begin{itemize}
 \item We have made a high-resolution 3.6\,cm imaging observation of a hard X-ray emission source
       (TKH\,8) associated with a protostellar clump (MMS\,2) in OMC-3 with the VLA. Two VLA
       sources were detected, both of which have the NIR counterparts.
 \item The centimeter emission is concluded to be free-free emission
       produced by a protostellar jet based on (1)~the association with class~I
       protostars, (2)~the low-mass nature of these protostars, (3)~the relation between
       the centimeter flux density and the momentum rate of the outflow, and (4)~the
       elongated structure aligned with the global outflow.
 \item The hard X-ray source is located $\sim$\,1--2\arcsec\ offset in the direction of
       the protostellar jet and outflow from the NIR sources. The origin of the hard
       X-ray emission was discussed in the framework of jet-induced shocks. The X-ray
       and the centimeter observations give independent constraints on the shock
       parameters, which are found to be consistent with each other.
\end{itemize}

\bigskip
We would like to thank Tae-Soo Pyo for updated information on fast outflow
observations. M.\,T. and N.\,K. express gratitude for the hospitality of NRAO staff in
the Array Operation Center in Socorro, New Mexico, USA. The National Radio Astronomy
Observatory is a facility of the National Science Foundation operated under cooperative
agreement by Associated Universities, Inc. M.\,T. is financially supported by Japan
Society for the Promotion of Science. K.\,K. is supported by a Grant-in-Aid for the 21st
Century COE ``Center for Diversity and Universality in Physics''.


\end{document}